# Forbidden Fe$^+$ Emission From Supernovae Remnants in M33

S.L. Lumsden[1,2] and P.J. Puxley[3]
[1] Department of Physics, University of Oxford, Keble Road, Oxford, UK
[2] Anglo-Australian Observatory, PO Box 296, Epping, NSW 2121, Australia * – sll@aaoepp.aao.gov.au
[3] Royal Observatory Edinburgh, Blackford Hill, Edinburgh, EH9 3HJ, UK – pjp@roe.ac.uk

22 April 1995

**ABSTRACT**

Supernovae remnants are known to be luminous sources of infrared [FeII] emission. By studying how the luminosity scales with age, environment and other relevant factors, we can construct an [FeII] luminosity function for supernovae remnants. This will enable us to predict supernovae rates in starburst galaxies that are too distant for individual remnants to be resolved. First, however, we require accurate luminosities for a sample of remnants of varying ages, and in varying physical environments. As part of this project we have carried out an initial study of a small sample of evolved (ages greater than a few thousand years) remnants in M33. From these data we tentatively conclude that there is evidence for the peak luminosity in the [FeII] lines of these sources to arise in a narrow range of ages. In other respects, the M33 remnants are similar to their galactic and Magellanic Cloud counterparts in the observed peak luminosity. From this, and internal evidence as to the environment present in these regions, we conclude that the luminosity of evolved remnants is only marginally dependent on density and metallicity.

## 1 INTRODUCTION

It has been realised for some time now that the intense [FeII] emission observed in the near infrared in starburst galaxies is likely to be due to supernovae remnants (SNR). For example, Moorwood & Oliva (1988) detected 1.644$\mu$m [FeII] emission from about half their sample of galaxies classed as star forming. More recent work has only confirmed this: in their study of compact sources in M82, Greenhouse et al. (1991), show that the observed 1.644$\mu$m to Br$\gamma$ ratio in the known radio SNR, 44.0+59.5 and 41.9+58.1, is greatly enhanced compared to the HII regions in the same galaxy. Similarly, Forbes et al. (1993) find a correspondence between some of the non-thermal compact radio continuum sources in NGC253 and enhanced [FeII] emission. This outline is also in agreement with infrared studies of SNR in our own galaxy, and in the Large Magellanic Cloud (eg. Oliva, Moorwood & Danziger 1989).

Some attempts have been made to use this fact to provide constraints on the nature of the initial mass function in starburst galaxies, most notably the work of Colina (1993). This parallels similar attempts to use high resolution radio data for the same purpose (e.g. Colina & Perez-Olea 1992) under the assumption that the non-thermal emission seen in luminous starburst galaxies arises in SNR. However, in principle, the use of [FeII] as a quantitative indicator of the supernova rate is superior to the radio technique, since the [FeII] emission from HII regions is very weak, whereas thermal radio emission from such sources can compete with non-thermal emission from SNR as the dominant source of radio luminosity in starburst systems (eg Condon et al. 1991). Additional uncertainty arises in the radio work since we cannot even be sure that the non-thermal emission seen does not arise in a low luminosity active nucleus (Lonsdale, Smith & Lonsdale 1993). However, in order to achieve our aim of using the [FeII] lines as diagnostic probes, we require a greater understanding of the evolution of the [FeII] line luminosity as a function of age in SNR in differing environments. For example, the brightest compact SNR's observed in M82 have 1.257$\mu$m [FeII] luminosities of $\sim 10^{32}$W, whereas a typical bright remnant in the LMC has a luminosity of $\sim 10^{30}$W (Oliva et al. 1989). In order to derive useful supernovae rates from even more distant galaxies in which we cannot resolve the individual remnants, we must be able to predict a typical SNR luminosity given that we know its age and something about its environment. Given a supernovae rate we can then place constraints on the stellar initial mass function, since the luminous radio supernovae are typically type II (or Ib), and hence their progenitors are massive stars with initial masses $> 8$ M$_\odot$.

In order to calibrate this method we have therefore embarked on a series of observations of individual SNR. It is envisaged that the complete sample will span a wide range of [FeII] luminosity and environment. We require accurate distances in order to derive accurate luminosities. Since the distance to most galactic sources is poorly defined, we have excluded such sources from our sample. Although there are nearby starburst galaxies in which individual bright SNR have been resolved by radio observations, thus giving us access to the high luminosity end of our sample (eg. M82 Huang et al. 1994, NGC253 Ulvestaad & Antonucci 1991, NGC1808 Saskia et al. 1990), there are few galaxies which have been systematically surveyed for complete samples of SNR. Of these, the best candidates are the Magellanic Clouds (eg. Matthewson et al. 1983) and M33 (eg. Long et al. 1990). This paper reports on our initial study of a sample from this latter set.

* Present address

## 2 THE SAMPLE AND THE OBSERVATIONS

Our sample of objects was chosen from the confirmed SNR in the catalogue of Long et al. (1990). Since Oliva et al. found a correlation between the $1.644\mu$m [FeII] line and H$\beta$, we imposed a cut using the Long et al. H$\alpha$ surface brightnesses of $5\times10^{-19}$Wm$^{-2}$arcsec$^{-2}$. The limit in $1.257\mu$m [FeII] surface brightness is then $\sim$10% of this value (using the relationship between $1.644\mu$m [FeII] and H$\beta$, the fact that the $1.257\mu$m/$1.644\mu$m ratio is $\sim$ 1.35, and a standard case B recombination H$\alpha$/H$\beta$ ratio, though we note this latter may not be strictly valid). A second requirement was that the sources should be relatively compact given that we were obtaining spectroscopic data. We imposed a limiting diameter of $\sim$10arcsec ($\sim$35pc in table 3 of Long et al). The sample discussed in this paper is incomplete: the complete surface brightness limited sample derived from Long et al. is given in Table 1, and the ten SNR we actually observed are noted, together with the date of observation and integration time. Ideally for our purposes, we would have used a radio selected sample, chosen such that it spanned a wide range in radio luminosity. However, at the time of our observations such a catalogue was unavailable, though one has now been published (Duric et al. 1993). In practice, the sources we observed do span a reasonable range of radio luminosity as will be shown later.

All objects were observed with the common-user array spectrometer CGS4 on UKIRT (Mountain et al. 1990) on the nights of 13, 14 and 15 November 1992. Conditions were assessed as being photometric from the measurements of the flux standards described below. CGS4 utilises a SBRC InSb 62$\times$58 array, sensitive from 1–5$\mu$m. The spatial resolution was 3$''$ per pixel, and the slit-width was also 3$''$. In order to obtain fully sampled data, the array is stepped 4 times over 2 pixels: this also enables single bad pixels to be removed from the data (Puxley, Beard & Ramsay 1992). The central wavelength for all observations was 1.26$\mu$m. We chose to observe the $1.257\mu$m a$^6$D$_{9/2}$–a$^4$D$_{7/2}$ [FeII] line since it is inherently stronger by a factor of 1.35 (Nussbaumer & Storey 1988) than the $1.644\mu$m a$^4$F$_{9/2}$–a$^4$D$_{7/2}$ line, allows simultaneous observation of Pa$\beta$, is significantly less affected by night sky emission bands than the $H$ window (though a strong and variable oxygen airglow line is centred at $1.27\mu$m – eg Ramsay, Mountain & Geballe 1992) and extinction is not a problem given the inclination of 55$^\circ$ of M33. The objects were observed in 2nd order, with a resolution of approximately 400.

As atmospheric standards we used the late-type A stars SAO54655 (A0), BS553 (A5V) and HD3029 (A3), from which the Pa$\beta$ absorption feature was removed by fitting a Lorenzian absorption profile. All standards were observed at similar air mass to the target SNR, eliminating the need to make airmass correction in our flux calibration. We observed one of these standards once every 1–2 hours. BS553 and HD3029 are also flux standards in the UKIRT standards list, with magnitudes of $J = 2.50$ and $J = 7.263$ respectively. Using these values we obtain $J = 5.17\pm0.05$ for SAO54655. Wavelength calibration was by means of krypton and xenon arc lamps. Comparison of the fluxes of the standard stars from repeated observations over the three nights provides a measure of the accuracy of the overall photometry. From this we estimate that the photometric errors on any given single spectrum are <10%.

The slit was oriented north-south for all observations. We pointed at the positions given by Long et al., though for DDB-2 we also searched near the location given by D'Odorico, Dopita & Benvenuti (1980). In all cases in which we detected emission, the peak was found to be within approximately 1–2 arcsec of the quoted positions. The objects observed are discussed more fully below. For all but DDB-2 and 013110+30210 we are confident that we observed at the location of the optical SNR. For those objects that were known from the optical data to be extended, we stepped the telescope a full slit width across the object until no flux was detectable in the same integration time as that for the central position (generally 320 or 640 seconds – see Table 1). We did not map further in those cases where we failed to detect [FeII] emission in the central position. In most cases we also integrated on the central position for a further considerably longer time in order to obtain good signal-to-noise on the less luminous SNR and to detect further weaker lines in the brighter SNR. These repeat observations also allowed us to check the accuracy of our photometry. For the compact sources the differences are less than 30%, for the more extended ones it is closer to 100%. We conclude therefore that our overall photometry for all sources is accurate to this level. Future surveys of these objects seeking accurate photometry should ideally use narrow band imaging techniques.

The reduction procedure for these data followed a standard procedure. First, the sky frames were subtracted from the corresponding object frame. It was found in practice that up to 10% of the original sky line flux was left after this subtraction. Therefore, a polynomial fit to the residual sky emission lines in this object–sky pair was then made, and the resulting fit removed to correct for this. The individual rows containing the positive and negative beams were then extracted. The data were then coadded, and corrected for atmospheric transmission by division by a standard star. Lastly spectra were then corrected for the assumed blackbody shape of the standard. Effective temperatures for the standards were assumed to be 8200K, 8720K and 9520K for BS553, HD3029 and SAO54655 respectively.

No extinction correction has been applied to this data. From Blair & Kirshner (1985), it can be seen that the extinction at $J$ is likely to be small and hence is entirely negligible for the purposes of calculating line ratios. For absolute line luminosities, the correction is mostly <10%. The observed fluxes are given in Table 2 and the spectra of those SNR with positive detections are shown in Figure 1. All spectra shown are of the single long exposures on the peak positions only. The error bars shown represent both overall photometric error on the spectrum, but also provide an indication of those regions where the sky transmission is less, or where the residual contamination from sky emission lines is greater. Since the positions observed away from the peak of the remnants were often of low signal-to-noise, weaker lines were not generally detected in these locations. Generally the line strengths listed in Table 2 will reflect integrated intensities for the strong lines but only peak intensities for the weaker ones, and comparison of ratios of these is not strictly valid. Therefore, the comparison of line ratios in what follows (especially section 3.2) applies to lines measured in the peak position alone.

## 3 RESULTS

### 3.1 Comments on Individual Remnants

*DDB-2:* The centroid of this object as given by Long et al. is ∼5arcsec east of the coordinates given by Duric et al. for the radio centroid. D'Odorico et al. (1980) give a position some 23arcsec south of the Long et al. one (this is the largest discrepancy between these two references for objects that are common to both that we actually observed). We tried both the Long et al. and D'Odorico et al. positions without detecting any [FeII] emission. However, since we did not map beyond these two central positions it is possible that we narrowly missed the object when observing the Long et al. centroid (the object itself is compact with a diameter of <4arcsec).

*DDB-4:* Weak [FeII] emission was detected on the central position. No emission was evident either 3arcsec east or west. There is a very small offset between radio and optical position well within the claimed accuracy of the Long et al. coordinates (2arcsec). Paβ was detected at ∼ 50% of the 1.257µm [FeII] flux.

*013022+30244:* No emission was evident. Again, the centroid given by Long et al. is some 5arcsec east of the radio centroid. The 20cm radio map of Duric et al. also shows considerable emission south-west of this location, though only very diffuse nebular emission appears to be present there. Again, it is possible that we may have missed this very compact optical source.

*DDB-6*: Very small offset between radio and optical position of 2arcsec, though in the same sense as above (radio position west of optical). [FeII] emission is extended: we found detectable emission in 3 adjacent slit positions. Very weak Paβ emission was also evident.

*DDB-7:* Radio and optical positions agree within the errors. Faint emission evident beyond the central slit position, but bright core of object is small. Weak Paβ emission evident.

*DDB-8:* Radio and optical positions agree within the errors, though we find peak of [FeII] emission to be in the slit position 3arcsec west of this. Emission is more diffuse than in the case of DDB-7. Paβ in this source is stronger at ∼30% of [FeII] on the core, and brighter than the [FeII] line in the west. This HI emission may be background.

*DDB-9:* Again we find [FeII] peaks to the west of the radio/optical centroids. Long et al. note that the object appears circular with the western edge being brighter. Moderately strong Paβ emission across the SNR.

*013109+30225:* Not detected in the radio survey of Duric et al. Faintest detection in our survey. No evidence for extended emission along the slit on the central position, so only one map position was observed.

*013110+30210:* Again no detected radio emission. No detectable [FeII] emission though we do detect the star seen to the south of the optical source. We are therefore confident that our positioning for this object is correct.

*DDB-11:* We find the peak to lie to the west of the radio/optical centroid. Long et al. note that the [OIII] emission also peaks north-west of the centroid. Weak Paβ to the east of the core, none on the brightest [FeII] spot.

For the three non-detections we can be absolutely confident of our positioning for only one (013110+30210). It is also worth noting that the compactness of the [FeII] emission does not necessarily follow the radio, or the quoted optical diameters from Long et al. We shall return to this issue in section 3.4.

### 3.2 Physical Conditions Within the SNR

We can use the ratios of the observed [FeII] lines to place constraints on the physical conditions present within the Fe$^+$ emitting zone. The atomic data of Nussbaumer & Storey (1988) and Pradhan & Zhang (1993) were used to interpret these ratios. The much larger collision strengths reported by Pradhan & Zhang compared to Nussbaumer & Storey (1980) reflects the fact that their computations take account of the many resonances that exist in the excitation of Fe$^+$ more accurately. Since the method employed should in principle be more accurate than that used in obtaining the older Nussbaumer & Storey (1980) data we will use their data in this analysis.

The 1.257µm line luminosities of the brighter remnants are $> 10^{29}$ W. For typical densities and temperatures, the luminosity of all the low lying [FeII] lines together is about ten times larger than this. Since a supernovae injects around $10^{44}$ J into the interstellar medium, if the typical [FeII] emitting time is only of the order of $10^4$ years (see below) then it is easy to see that Fe$^+$ alone is not an important coolant.

All of the [FeII] lines detected by us in these observations arise within the a$^4$D–a$^6$D multiplet (see Nussbaumer & Storey 1988 for a complete list of the lines in this multiplet). All of our detections are consistent within the errors with the available atomic data, as are the upper limits on other lines in this mulitplet, with the exception of the 1.279µm a$^4$D$_{3/2}$–a$^6$D$_{3/2}$ transition. This line is blended with Paβ in many cases, and also suffers from being coincident with strong night sky lines of OH and O$_2$. We therefore attribute the discrepancy in derived electron density using this line to our inability to both fit the line accurately and to residual sky contamination. Therefore, all electron density determinations are derived using the ratio of the 1.294µm a$^4$D$_{5/2}$–a$^6$D$_{5/2}$ to 1.257µm a$^4$D$_{7/2}$–a$^6$D$_{9/2}$ transitions. Limits are derived by measuring the flux in a 2 pixel wide segment at the correct wavelength for the 1.294µm line.

Both Oliva, Moorwood & Danziger 1990 (for the SNR RCW 103) and Rudy et al. 1992 (for the emission line star MWC 922), reach substantially the same conclusions on the comparison of observations and theory using the older collision strengths of Nussbaumer & Storey (1980). Since the new strengths of Pradhan & Zhang are substantially larger in most cases, the similarity of these conclusions indicates that most of the changes are due to an overall scaling of all the collision strengths (see the discussion in Oliva et al. 1990). To see the effect the changes have had, we consider also the electron densities as derived from the optical data of Blair & Kirshner (1985) and Smith et al. (1993). The [SII] transitions at 6717Å and 6731Å are sensitive to densities in the range $n_e = 100 - 10^4 \text{cm}^{-3}$. Table 3 shows the densities derived using this method compared to the densities derived using the [FeII] ratios. The error in the derived $n_e$ comes from the effect of varying the electron temperature (a negligible effect for the [FeII] densities). Since we have observed the brightest objects in the Long et al. catalogue, the observed signal-to-noise in the optical lines is always large.

From these results we can see that the derived electron densities in general are less from the [SII] ratios than from the [FeII] ratios. This fact was first noted by Oliva, Moorwood & Danziger (1989) for a small sample of galactic and LMC SNR using the older Nussbaumer & Storey (1980) atomic data. However, we see a smaller offset than they did. We therefore also checked the results of Oliva et al. using the newer atomic data for $Fe^+$ as well. They used the density sensitive ratio 1.599$\mu$m $a^4D_{3/2}$–$a^4F_{7/2}$ to 1.644$\mu$m $a^4D_{7/2}$–$a^4F_{9/2}$ ratio to derive electron densities in a small sample of SNR in our galaxy and the Large Magellanic Cloud. Using the Pradhan & Zhang collision cross-sections, the predicted values of $n_e$ are all less than given by Oliva et al. These values are also given in Table 3. This reduces the discrepancy between the inferred electron densities from the [FeII] and [SII] ratios they see, though it does not eliminate it, just as is the case for our observations of the M33 SNR. Given the change that has occurred in moving from the Nussbaumer & Storey (1980) data to the Pradhan & Zhang data we must caution that the remaining discrepancies may also still be due to inaccuracies in the collision cross sections. However, their rationale for this discrepancy, the low critical density for the [SII] lines (and in particular the 6717Å line), is also perfectly valid. It would be extremely valuable to obtain further data on IR transitions of [FeII] in bright galactic remnants with the current generation of IR spectrometers to further test the reliability of the atomic data.

Lastly, we note that there is a relatively wide span in SNR [FeII] luminosity in our sample, but little evidence for there being a wide range in density. We therefore conclude that the ambient density plays only a minor role in determining the [FeII] luminosity of the SNR in our sample.

### 3.3 Comparison of [FeII] and other observed emission

Both Smith et al. (1993) and Blair & Kirshner (1985) have obtained optical spectra of the SNR in M33 in the wavelength range ~ 4000Å–7500Å. It is worthwhile comparing the fluxes arising from different species, to seek possible correlations. It has already been noted (for example Smith et al.) that the forbidden line fluxes tend to be correlated if they are not strongly dependent on the physical conditions in the SNR. Unfortunately, Blair & Kirshner observed only some of those objects in the D'Odorico et al. sample. We therefore compare our data with that of Smith et al. who observed all the SNR under consideration here. Smith et al. only observed at one slit position per object, but our own data suggests that the sources are sufficiently compact that the bulk of the emission does lie in a narrow slit. Smith et al. also aligned their slit over the brightest regions of the SNR: the line fluxes as measured by them for the objects in our sample are given in Table 4. Also given there are optical sizes from Long et al. (1990), radio sizes and fluxes from Duric et al. (1993) and velocity information from from Blair, Chu & Kennicutt 1988 and Blair & Davidsen 1993.

In Figures 2a-d we show how the 1.257$\mu$m [FeII] flux correlates with 6300Å [OI], 6731Å [SII], 6365Å H$\alpha$ fluxes and 6cm radio flux, and present linear correlation coefficients for all these tests in Table 5. In this analysis, objects with only upper limits have been excluded. The 6731Å [SII] line was chosen since it has a much higher critical density than the 6717Å line. Clearly there is a well defined correlation between [FeII] emission and the optical lines, but only a weak correlation between the forbidden line fluxes and the measured radio continuum. The discrepancies that exist between our [FeII] data and the optical emission line data are probably due largely to the different effective apertures being used.

Most of these correlations (or lack of them) can be understood in terms of the source of the emission. Generally, the optical and infrared line emission in SNR arises in radiative shocks, whereas the radio emission arises from synchrotron radiation. In particular, the fluxes of [SII], [FeII] and [OI] should all be strongly correlated since they all arise in a region in which hydrogen is largely neutral. This is true both of shock excited emission, and of photoionisation (as is seen in the Crab Nebula for example - e.g. Fesen & Kirshner 1982, Henry & MacAlpine 1982). By contrast, [OII], with a similar ionisation potential to hydrogen, comes from a zone in which hydrogen is at least partially ionised. In terms of shock models, the various regions that we can observe emission from are the immediate post-shock region (where hydrogen is fully ionised, [OIII] emission arises and iron if present is in a higher ionisation state), a transition region in which gas recombines rapidly (the [OII] region) and a larger volume of gas in which ionised stages of iron are rapidly converted into $Fe^+$ by charge exchange reactions with the now dominant neutral hydrogen (see, for example, Oliva et al., Van der Werf et al. 1993).

The one extremely discrepant point in Figure 2d, DDB-9, may have a different nature from the other remnants, since it is also a bright x-ray source (eg Long et al. 1990: DDB-7 has also been detected in x-rays but shows no signs of being different from the other sources in other regards) However, from our current data, we cannot say anything further.

### 3.4 Correlation of [FeII] luminosity and SNR properties

In trying to determine how the [FeII] luminosity correlates with the age of the remnant, we must first decide on a suitable observable quantity that defines this latter parameter. In principle, virtually all of the remnants in our sample should be either in an adiabatic or radiative expansion phase. Long et al. (1990) found that $N(<D) \propto D^{2.1}$ (number of SNR with diameter less than $D$pc) for M33, which should be compared with an exponent of 2.5 for adiabatic expansion and 1 for free expansion. Hence, we will assume that the standard Sedov solution can be applied to the evolution of these remnants (see, eg. Lozinskaya 1992 for a fuller description of the evolution of SNR). Although this is unlikely to be accurate at early times, it should provide a reasonable approximation at these late epochs. We have used this solution to derive approximate ages, in the first instance by using the relationship between age and remnant diameter, but also by considering the relationship between shock velocity and age as well. According to the Sedov solution, these quantities are then related by

$$R_s \propto (E_0/n_0)^{0.2} t^{0.4} \qquad v_s = 0.4 R_s/t,$$

where $R_s$ and $v_s$ are the radius and velocity of the shock, $t$ is the remnant age, $E_0$ the initial energy of the supernova and $n_0$ the density of the amibient medium. For typical parameters ($E_0 = 10^{44}$ J, $n_0 = 1$cm$^{-3}$), $R_s \sim 0.34 t_{\rm yr}^{0.4}$ pc.

Although it is likely that the remnants as a whole are still undergoing adiabatic expansion, radiative shocks are prominent in these objects. One possibility is that the [FeII] lines reach maximum luminosity when the remnant as a whole undergoes the transition from adiabatic to radiative expansion. Detailed models (eg. Cioffi & McKee 1988) show this to be a relatively slow process, and indeed predict the SNR luminosity to peak just as the radiative phase is reached. Typically, this would occur after $\sim 10^5$ years in their models, rather older than the brightest remnants that we see.

There are many different estimates of the sizes of these SNR. Long et al. give optical diameters based on the geometric mean of the dimensions of the SNR found in their optical images. Duric et al. quote diameters based on the deconvolution of their radio data. These do not always agree as shown by Table 4. Blair & Davidsen (1993) find different results to either of these groups based on analysis of Hubble Space telescope imaging data of DDB-2 and DDB-8. Blair, Chu & Kennicutt (1988) also give different diameters as determined from their echelle spectroscopy. Similarly, higher resolution VLA data (Goss & Viallefond 1985) shows that DDB-7 is substantially smaller than either Long et al. or Duric et al. report. We also find a smaller diameter for this source ($\sim 3.5''$) from an [FeII] image obtained with IRCAM3 at UKIRT.

Given these problems with determining the 'true' diameters of these SNR, it is not altogether surprising that the relation between [FeII] flux and optical diameter shown in Figure 3a is so poor (using radio diameters is no better as the correlation coefficients in Table 5 show). We therefore rule out using diameter as an accurate measure of age. The better correlation found between the optical emission lines and the optical radii is effectively a selection effect. Smith et al. (1993) align their slit to 'include the brightest portions of the objects' from the original Long et al. images. It is not surprising then that the measured flux correlates with the original data, but it does not discount the possibility raised by Blair & Davidsen (1993) that the measured diameters of all these sources are wrong.

By comparison, in Figure 3b we show the correlation between [FeII] flux and velocity for the remnants. The correlation here is much stronger (Table 5), with the expected trend that the most luminous sources have the broadest H$\alpha$ lines. For typical values of the parameters (initial energy, density of the interstellar medium), the Sedov solution (eg. Lozinskaya 1992) implies that the brightest remnants in this sample are $\sim 10^4$ years old. We also note the poor correlation between optical radius and velocity width (Figure 3c): the measured values of the diameters and velocity widths for the two most [FeII] luminous remnants in our survey are completely incompatible with the Sedov solution. These discrepancies are not the result of obvious density differences, since the data from the [SII] lines shows that most of these SNR have similar densities. Hence our conclusion on the reliability of the measured diameters as an age probe is borne out by this comparison as well. Unfortunately, since we do not have measured velocity widths for any of the most compact optical remnants, we cannot tell if they are also genuinely young using this measure of age.

Therefore, from this dataset alone, we cannot rule out either of the two possibilities: first, that the forbidden line luminosity is largest for the youngest remnants with the most energetic shocks; secondly, that the luminosity peaks at the point which most likely represents the slow transition from adiabatic to radiative expansion. We can say from the observed weakness of [FeII] emission in many of the younger remnants in our galaxy and the LMC, that there must be some lower cut-off as well as a higher one (Hester et al. 1990 find a 1.644$\mu$m luminosity for the Crab remnant $\sim 10^{27}$ W, and we similarly find little evidence of [FeII] emission in the youngest of the LMC remnants we have studied). Hence the idea that [FeII] luminosity may peak sharply with the age of the remnant is consistent with our results but further observational data is required to confirm it.

## 4 DISCUSSION

From our incomplete sample of SNR in M33, we have demonstrated the fundamental similarities between the infrared [FeII] line strengths and those of well studied optical transitions with equivalent ionisation potential such as [SII]. Therefore, [FeII] lines are good tracers of SNR properties. This is especially true for regions of high extinction (given their longer wavelength) and where HII regions coexist in the beam with SNR (since [FeII] emission from HII regions is essentially negligible).

It is useful to compare the typical luminosities that we find here with the values found for the 1.644$\mu$m line by other groups. The only well studied older remnants are in our own galaxy and the LMC (see Oliva et al. and references therein). Typical bright remnants there have equivalent 1.257$\mu$m luminosities of $\sim 10^{30}$ W, in reasonable agreement with the values we find for M33. These peak luminosities also appear to arise from remnants which are typically a few thousand years old. In a future paper we will present our own findings on the [FeII] luminosities and structures of a larger sample of LMC remnants. It seems likely from the work completed to date that the dominant factor in determining the luminosi-ties for these evolved remnants is their age. It is unlikely that metallicity is the cause (both the LMC and M33 have similar metallcities, yet the brightest [FeII] source in the LMC, N49, is almost ten times brighter than DDB-8). Similarly, variations are seen in the density of the bright remnants (as is the case here for DDB-7 and DDB-8) so again this is unlikely to be a major factor. The strong correlation that we see between luminosity and bulk velocity indicates that age is the dominant mechanism within a galaxy, and this is likely to be true when comparing different systems as well.

Given these conclusions, it is worth discussing briefly the nature of the compact radio sources seen in many nearby starburst galaxies, and commonly assumed to be SNR (though see Ulvestaad & Antonucci 1994). The compact sizes of these 'remnants' naturally leads to the assumption of a young age for them. However, there are two problems with this simple interpretation. First, the narrow width of the [FeII] emission detected by Greenhouse et al. in M82 is rather puzzling if this is true. Secondly, the radio data shows no strong evidence for a rapid fading of these sources as is seen in isolated radio supernovae (eg. Ulvestaad & Antonucci 1994, Huang et al. 1994). If we assume that these SNR are young, then we must further conclude that the emission mechanism in these remnants is a different process to the radiative cooling of filaments that we see in these older remnants. One possibility has received considerable attention to model the luminous radio emission seen from some extragalactic SNR. In this model the supernova blast wave overtakes a dense circumstellar shell, and it is largely the excitation of this shell that we see at early times (eg. Chevalier & Frannson 1994). In the high pressure environment in the nuclei of starburst galaxies, it seems likely that such shells would be confined even closer to the progenitor star, and hence may lead to even greater emission. Even this model has some difficulties however, since it predicts broad emission lines should be present. Further study of the properties of remnants in a high pressure environment, both theoretical and observational, is highly desirable.

### Ackowledgments

We would like to thank Joel Aycock for his customary efficiency at the controls of UKIRT. SLL acknowledges support from the SERC through a Postdoctoral Fellowship whilst part of this work was carried out.

| Name | Diameter (arcsec) | $\sigma$(H$\alpha$) | Date(s) of Observation | Integration Time (s) |
|---|---|---|---|---|
| DDB-2 | 3.6 | 4 | 13 & 14 Nov 1992 | 3×640 |
| DDB-4 | 6.9 | 5 | 15 Nov 1992 | 3520 + 2×320 |
| 013022+30244 | 3.3 | 1 | 14 Nov 1992 | 2560 |
| DDB-6 | 3 | 7 | 13 & 15 Nov 1992 | 1920 + 3×320 |
| DDB-7 | 9.7 | 2 | 15 Nov 1992 | 2560 + 3×640 |
| DDB-8 | 8.3 | 7 | 13 & 15 Nov 1992 | 2080 + 4×320 |
| 013049+30270 | 5.7 | 0.5 | not observed | |
| 013059+30177 | 8 | 0.9 | not observed | |
| DDB-9 | 4.8 | 4 | 13 & 15 Nov 1992 | 1440+2×320 |
| 013109+30225 | 4.3 | 1 | 14 Nov 1992 | 3360 |
| 013109+30168 | 5.1 | 0.9 | not observed | |
| 013110+30210 | 1.8 | 3 | 13 Nov 1992 | 960 |
| 013110+30182 | 5.1 | 0.5 | not observed | |
| DDB-11 | 5.4 | 2 | 14 & 15 Nov 1992 | 1920+2×320 |
| 013125+30192 | 8.3 | 1 | not observed | |
| DDB-14 | 2.9 | 2 | not observed | |
| DDB-15 | 6 | 1 | not observed | |

**Table 1.** The complete diameter and surface brightness limited sample of SNR in M33. The H$\alpha$ surface brightness is given in units of $10^{-15}$Wm$^{-2}$arcsec$^{-2}$. Diameters and surface brightness values are taken from Long et al. (1990)

| Line | $\lambda$ ($\mu$m) | DDB-4 | 013022 +30244 | DDB-6 | DDB-7 | DDB-8 | DDB-9 | 013109 +30225 | 013110 +30210 | DDB-11 |
|---|---|---|---|---|---|---|---|---|---|---|
| [FeII] | 1.257 | 2.3 ± 0.3* | < 0.35 | 32.4 ± 1.3 | 35.7 ± 1.2 | 47.2 ± 1.8 | 24.8 ± 1.1 | 0.8 ± 0.12 | <1.5 | 18.7 ± 1.1 |
| [FeII]? | 1.271 | | | | 0.92 ± 0.34* | | | | | |
| [FeII] | 1.279 | | | 1.2 ± 0.5 | 2.9 ± 0.8* | 1.3 ± 0.7* | | | | |
| Pa$\beta$ | 1.282 | 1.3 ± 0.4* | | 4.7 ± 2.0 | 6.8 ± 1.2 | 28.6 ± 3.7 | 8.5 ± 1.3 | | | 4.8 ± 1.3* |
| [FeII] | 1.294 | | | 2.3 ± 0.7* | 2.5 ± 0.4* | 1.2 ± 0.4* | | | | |
| [FeII] | 1.321 | 0.86 ± 0.27* | | 8.3 ± 1.3 | 9.4 ± 0.8 | 11.3 ± 1.2 | 5.1 ± 0.8 | | | 2.9 ± 0.7* |
| [FeII] | 1.328 | | | | 0.93 ± 0.23* | | | | | |

**Table 2.** Observed line fluxes seen in the M33 SNR. No extinction corrections have been applied. Lines marked with an * were only detected at the peak of the emission, as discussed in the text. All fluxes are in units of $10^{-18}$Wm$^{-2}$.

| Object | $n_e$ (cm$^{-3}$) [FeII] | $n_e$ (cm$^{-3}$) [SII] |
|---|---|---|
| DDB-4 | <$10^4$ | 250±100 |
| 013022+30244 | | 375±200 |
| DDB-6 | 1000±700 | 500±250 |
| DDB-7 | 5000±2000 | 1500±500 |
| DDB-8 | 3000±2000 | 600±200 |
| DDB-9 | < 5000 | 450±200 |
| 013109+30225 | < $10^4$ | 375±200 |
| 013110+30210 | | 450±200 |
| DDB-11 | < 1000 | 650±200 |
| Data from Oliva et al. (1989) | | |
| RCW103 | 3000±1000 | 1250±350 |
| Kepler | 8000±3000 | >8000 |
| N63A | 4500±1500 | 2000±600 |
| N49 | 4000±1500 | 850±150 |
| N103B | 5000±3000 | >8000 |

**Table 3.** Derived electron densities from [SII] (Smith et al. 1993) and [FeII] transitions. The error bars for the [SII] data reflect the uncertainty in the electron temperature rather than any observational error. The observed signal to noise is large enough for the [SII] lines that this is valid. The variation with electron temperature for the [FeII] lines is negligible for most values of $n_e$. The electron densities for the Oliva et al. sample derived from the [SII] lines are taken from that source. The densities derived from the [FeII] lines have been calculated as described in the text.

| Object | [SII] | [OI] | Hα | Radius (opt) | Radius (radio) | Velocity | 6cm Flux |
|---|---|---|---|---|---|---|---|
| | ($\times 10^{-18}$Wm$^{-2}$) | | | (arcsec) | | (kms$^{-1}$) | (mJy) |
| DDB–2 | 4.1 | 2.0 | 11.5 | 3.6 | 3.4 | 525 | 0.3 |
| DDB–4 | 3.7 | 3.7 | 20.3 | 6.9 | 3.5 | 365 | 0.4 |
| 013022+30244 | 2.1 | 1.2 | 5.9 | 3.3 | 1.0 | | 0.1 |
| DDB–6 | 8.9 | 6.2 | 16.6 | 3.0 | 2.5 | 700 | 0.2 |
| DDB–7 | 37.5 | 16.3 | 72.9 | 9.7 | 5.2 | 660 | 0.8 |
| DDB–8 | 33.3 | 11.2 | 70.7 | 8.3 | 5.6 | 625 | 1.1 |
| DDB–9 | 5.7 | 2.4 | 14.5 | 4.8 | 4.6 | 540 | 2.4 |
| 013109+30225 | 3.0 | 1.7 | 5.0 | 4.3 | | | < 0.2 |
| 013110+30210 | 1.7 | 1.0 | 2.6 | 1.8 | | | < 0.1 |
| DDB–11 | 11.7 | 8.0 | 20.1 | 5.4 | | 370 | 0.2 |

**Table 4.** Published values for line fluxes, sizes, velocity data and radio fluxes taken from Smith et al. 1993, Long et al. 1990, Duric et al. 1993, Blair, Chu & Kennicutt 1988 and Blair & Davidsen 1993 as noted in the text. Limits are given where appropriate, and a blank is left where no published data exists.

| | [SII] | [OI] | Hα | Radius (opt) | Radius (radio) | Velocity | 6cm Flux |
|---|---|---|---|---|---|---|---|
| [FeII] | 0.79 | 0.73 | 0.75 | 0.39 | 0.51 | 0.83 | 0.22 |
| [SII] | | 0.81 | 0.98 | 0.84 | 0.74 | 0.52 | 0.18 |
| [OI] | | | 0.93 | 0.82 | 0.67 | 0.46 | 0.04 |
| Hα | | | | 0.90 | 0.78 | 0.47 | 0.21 |
| Radius (opt) | | | | | 0.79 | 0.07 | 0.25 |
| Radius (radio) | | | | | | 0.16 | 0.62 |
| Velocity | | | | | | | 0.17 |

**Table 5.** Linear correlation coefficients for the various measured parameters of the SNR discussed in the text.

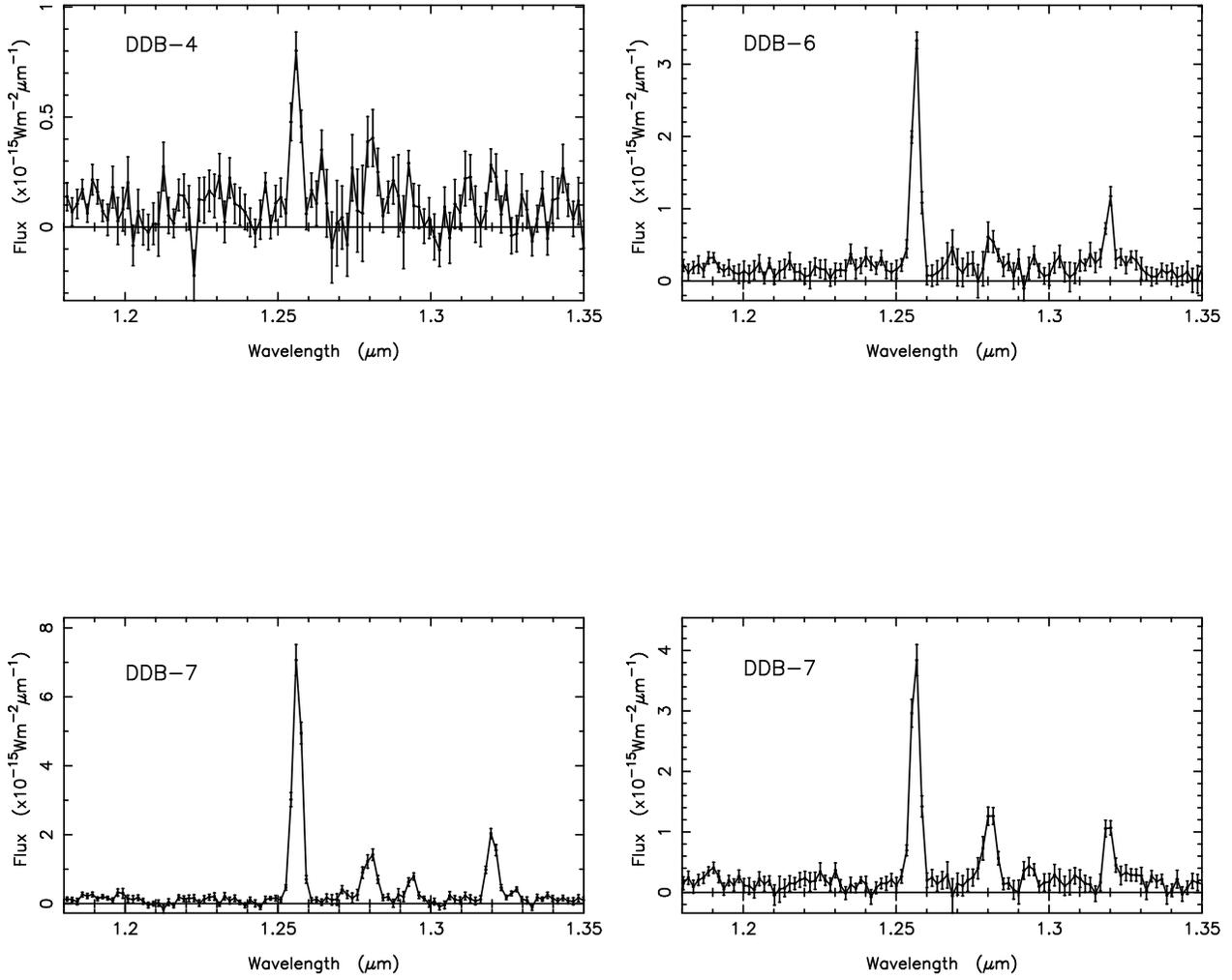

**Figure 1.** *J*–band spectra for DDB-4, DDB-6, DDB-7, DDB-8, DDB-9, DDB-11, 013109+30225. Spectra are presented for the peak position only.

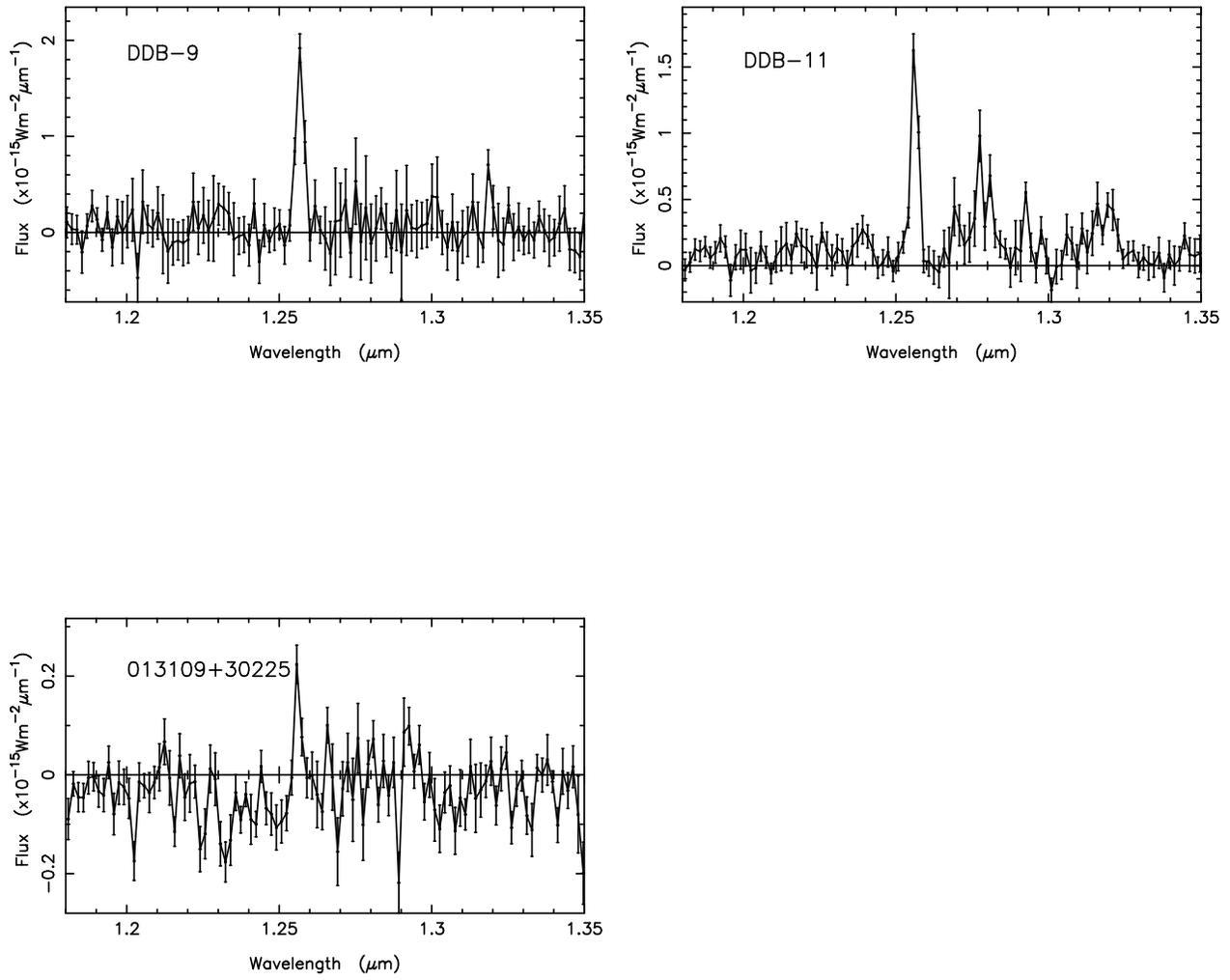

Figure 1.

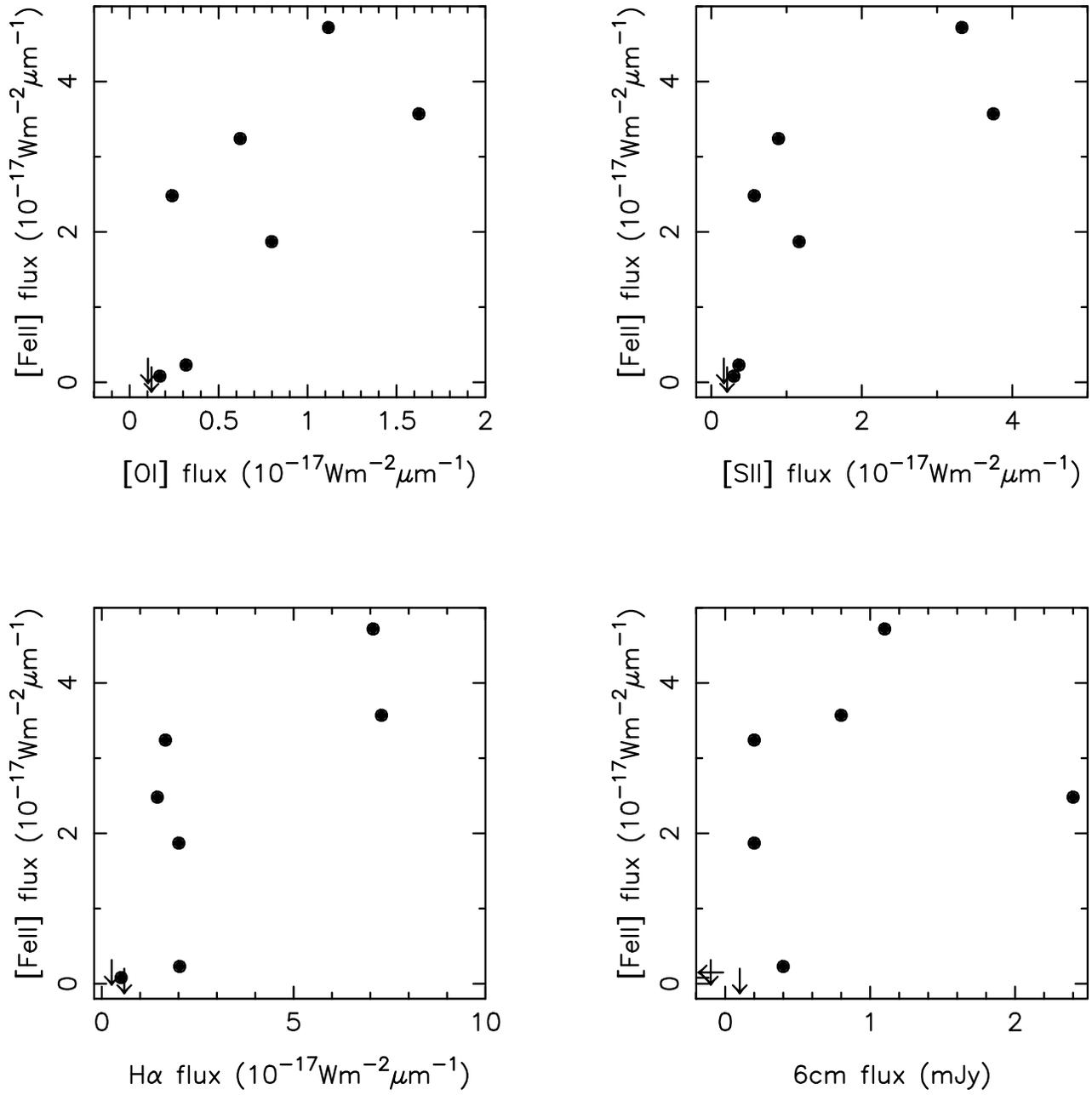

**Figure 2.** Scatter plots for 1.257μm [FeII] against (a) 6300Å [OI] (b) 6731Å [SII] (c) Hα and (d) 6cm radio flux density. Upper limits are shown as arrows, and are 3σ limits.

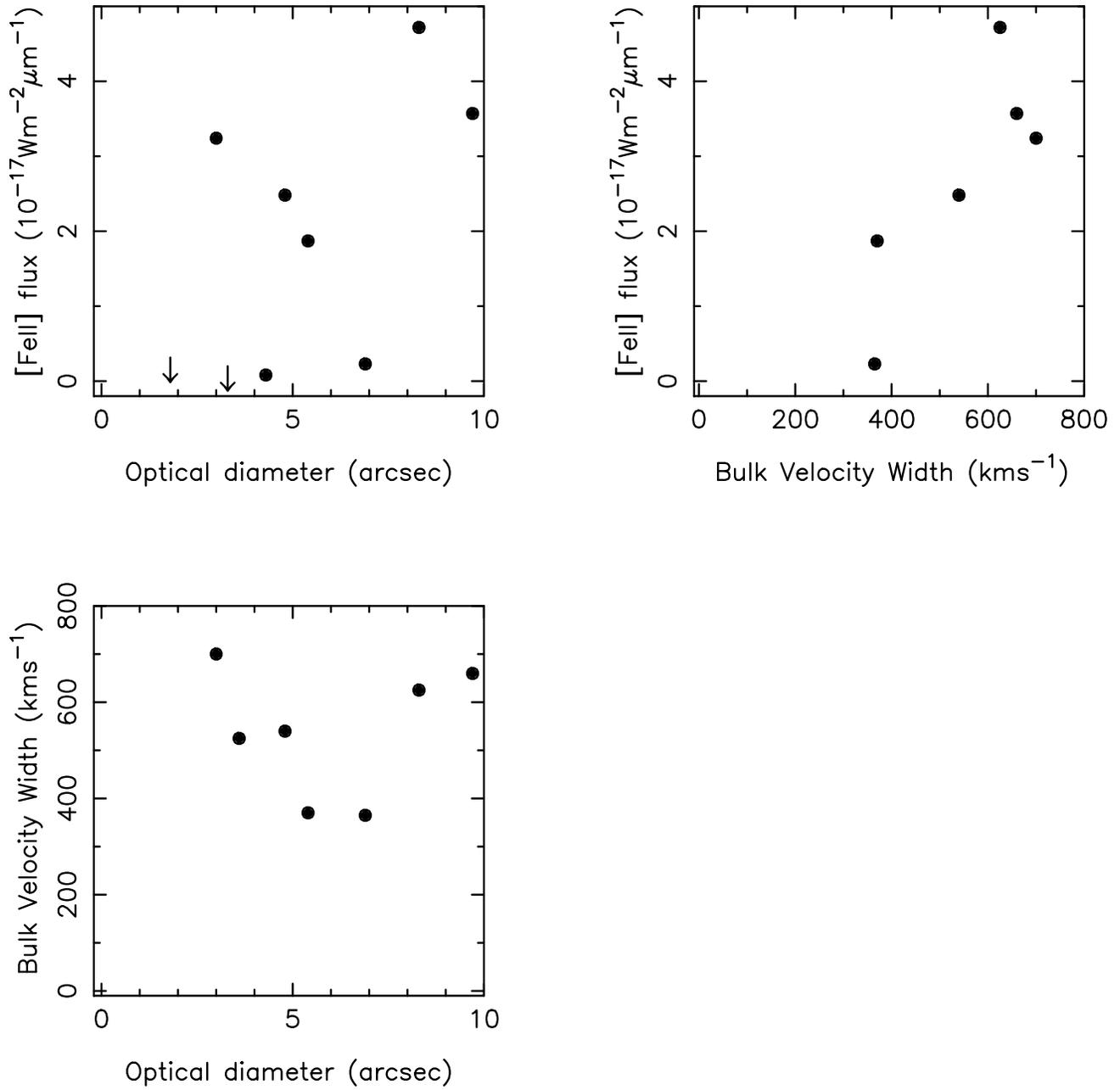

**Figure 3.** Scatter plots for 1.257$\mu$m [FeII] against (a) optical diameter and (b) bulk velocity width (see Blair et al. 1988 for a discussion of the velocity widths). Also shown is the relationship between velocity width and radius.